# Scale and curvature effects in principal geodesic analysis

Drew Lazar[a], Lizhen Lin[b]

[a]*Department of Mathematics, Ball State University*
[b]*Department of Applied and Computational Mathematics and Statistics, The University of Notre Dame*

**Abstract**

There is growing interest in using the close connection between differential geometry and statistics to model smooth manifold-valued data. In particular, much work has been done recently to generalize principal component analysis (PCA), the method of dimension reduction in linear spaces, to Riemannian manifolds. One such generalization is known as principal geodesic analysis (PGA). This paper, in a novel fashion, obtains Taylor expansions in scaling parameters introduced in the domain of objective functions in PGA. It is shown this technique not only leads to better closed-form approximations of PGA but also reveals the effects that scale, curvature and the distribution of data have on solutions to PGA and on their differences to first-order tangent space approximations. This approach should be able to be applied not only to PGA but also to other generalizations of PCA and more generally to other intrinsic statistics on Riemannian manifolds.

## 1. Introduction

Principal component analysis (PCA) is an important statistical method for dimension reduction and exploration of the variance structure of data in a linear space. PCA has been generalized to data in smooth manifolds in various principal geodesic procedures in which projections are done to explanatory submanifolds which serve as non-linear analogues of the linear subspaces of PCA.

*Principal geodesic analysis* (PGA), as introduced in [10], successively identifies orthogonal explanatory directions in the tangent space at the intrinsic mean of data and then exponentiates the span of the results to form explanatory submanifolds. In [10] first-order tangent space approximations of PGA were formulated. Subsequently methods for exact computation of PGA in specific manifolds were offered as in [17] and [27]. Then in [29], using the derivative of the exponential map and ODEs if necessary in gradient descent algorithms, procedures to find exact solutions in a general class of manifolds were outlined.

As pointed out in [28], however, exact computation of PGA can be computationally complex and time-intensive, and thus there is interest in determining the accuracy and effectiveness of first-order approximations to PGA. This will depend on the distribution of data and its dispersion from the tangent space, the curvature and shape of the manifold in question and the interaction of these factors.

For illustration, as in [28], consider the position of the "wrist" of a moving robotic arm while its "elbow" and "body" are fixed. In Figure 1 the motion is restricted to a two-dimensional surface. To analyze the movement of the wrist one might collect motion capture data as represented by the red dots in the figure. Formulating the surface as a Riemannian manifold and using intrinsic distances, an intrinsic mean of the data, $\mu$, might be located. Then a geodesic through $\mu$, represented by the blue curve on the surface, that best fits the data or best accounts for the data's variability might be identified.

One can find a linear direction of maximum variability, the unit vector $v_{1,0}$ in Figure 1, of data projected by the Riemannian log map to the tangent space at the intrinsic mean. $v_{1,0}$ will be an approximation of the unit vector tangent to the geodesic $v_1$. Generally the greater the local curvature of the surface the less accurate this approximation will be with scale of the data or its dispersion from the intrinsic mean augmenting this effect. Conversely, projections to the tangent space will converge to the data, intrinsic distances will converge to tangent space distances and $v_{1,0}$ will converge to $v_1$ as the data draws in towards $\mu$.

In this paper we quantify such effects by introducing scaling parameters on projections of data to the tangent space and by obtaining Taylor expansions of solutions to PGA procedures in these parameters. Leading terms, such as $v_{1,0}$ in Figure 1, will originate from the Euclidean structure in the tangent space. Next-order terms will demonstrate how local curvature and scale interact to contribute to differences between first-order approximations and exact solutions. This not only allows for more accurate closed-form approximations of PGA but should also contribute to a better understanding of the parts of PGA and corresponding statistics. In this paper data in three types of symmetric spaces which have regular application are considered. Also using [17, 27, 29] we can compute exact solutions in these spaces which allows for comparison and testing.

*1.1. Outline*

Section 2 includes notations and definitions. In Section 3 a proposition which allows the expansion of PGA directions in this paper is stated and proved. In

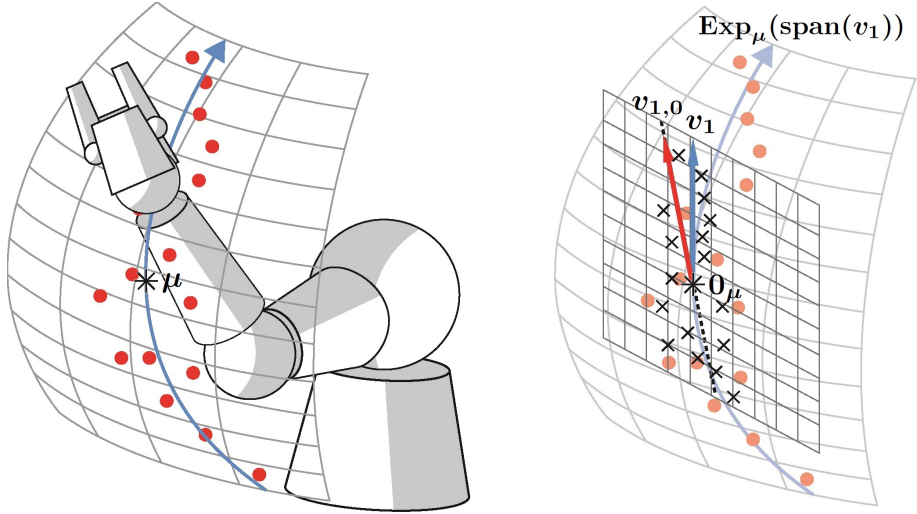

Figure 1: First PGA motion capture data

Section 4 we review the geometry of the $n$-spheres and obtain and test expansions using our proposition. We also carry out experiments on data sampled from an anisotropic log-normal distribution on the unit $n$-sphere to show improved approximations. In Section 5 we review the geometry of the space of positive definite matrices and obtain expansions using our proposition and computer algebra. In Section 6 we review the geometry of the special orthogonal group and obtain expansions of PGA in this space. Also, in Section 6.3 we take a closer look at PGA in Lie groups in [11] to show how expansions can give insight into the formulation of such intrinsic manifold statistics. In Section 7, using expansions, we obtain improvements of the *linear difference indicators* introduced in [28]. In Section 8 we discuss the results and consider their applications in similar contexts.

## 2. Notations and Definitions

Let $M$ be a Riemannian manifold with Riemannian metric $p \to \langle\ ,\ \rangle_p$ for $p \in M$. Given $p \in M$, $\mathrm{T}_p M$ is the *tangent space at $p$*. The unit sphere at $\mathrm{T}_p M$ is then $\mathrm{S}_p M = \{X \in \mathrm{T}_p M; \langle X, X\rangle_p = 1\}$. The *Riemannian exponential* and *Riemannian log maps*, are denoted by $\mathrm{Exp}_p : \mathrm{T}_p M \mapsto M$ and $\mathrm{Log}_p : M \mapsto \mathrm{T}_p M$, respectively. Given smooth manifolds $M_1$ and $M_2$, $p \in M_1$ and smooth mapping $\lambda : M_1 \to M_2$ we denote the *differential of $\lambda$ at $p$* by $d_p \lambda$. Then given smooth function $f : M \to \mathbb{R}$ and $p \in M$ *the gradient of $f$ at $p$* is denoted $\nabla_p f$ so that $\langle \nabla_p f, X \rangle_p = d_p f(X)$ for all $X \in \mathrm{T}_p M$. Differential geometry texts [6] and [25] provide a background for and definitions of these concepts.

All the manifolds we will deal with in the paper will be of the class defined below.

**Definition 1.** Let $M$ be a connected Riemannian manifold. $M$ is a *symmetric space* if and only if for every $p \in M$ there exists an isometry $\phi_p : M \circlearrowleft$, such that
$$\phi_p(p) = p \quad \text{and} \quad \forall_{X \in \mathrm{T}_p M} \ d_p \phi_p(X) = -X.$$

The tensors defined below are used to measure curvature in Riemmanian manifolds.

**Definition 2.** The *curvature tensor* is the $(1,3)$ tensor given as
$$\mathrm{R}(x,y)w = \nabla_x \nabla_y w - \nabla_y \nabla_x w - \nabla_{[x,y]} w$$
for vector fields $x, y, w$ where $\nabla$ is the covariant derivative and $[\cdot, \cdot]$ is the Lie bracket. The *Riemann curvature tensor*, also denoted by R, is the $(0,4)$ tensor given, for all $x, y, w, z \in \mathrm{T}_p M$, as
$$\mathrm{R}_p(x,y,w,z) = \langle \mathrm{R}(x,y)w, z \rangle_p.$$

The Riemann curvature tensor characterizes the more geometrical description of curvature given below.

**Definition 3.** The *sectional curvature* of $\sigma_{v,q} = \mathrm{span}(\{v, q\})$ is given, for all $v, q \in \mathrm{T}_p M$, as
$$K(\sigma_{v,q}) = \frac{\mathrm{R}_p(q, v, v, q)}{\langle q, q \rangle_p \langle v, v \rangle_p - \langle q, v \rangle_p^2}. \tag{1}$$

The sectional curvature is the Gaussian curvature of $\mathrm{Exp}_p(\sigma_{v,q})$. Also, as in [4, p. 16] the Riemann curvature tensor is completely characterized by the sectional curvature.

The following definition generalizes orthogonal projection in an inner product space to projection to submanifolds.

**Definition 4.** Let $V_k = \{v_1, \ldots, v_k\} \subset \mathrm{T}_\mu M$ and $H(V_k) = \mathrm{Exp}_\mu \{\mathrm{span}(V_k)\}$ with $\mu \in M$. For $p \in M$ the *projection* of $p$ to $H(V_k)$ is
$$\pi_{H(V_k)}(p) = \underset{x \in H(V_k)}{\mathrm{argmin}} \ \mathrm{d}(x, p).$$

As in [29], existence of projection is guaranteed when $H(V_k)$ is compact which will be the case when projecting in the special orthogonal group and in the $n$-spheres in this paper. Then as in [16] we will have uniqueness of projection almost everywhere in these manifolds. Throughout we will assume uniqueness of projection in the special orthogonal group and in the $n$-spheres. Also, with the positive definite matrices having non-positive curvature, as in [10] there is existence and uniqueness of projection for every $H(V_k)$.

The first intrinsic statistic we formulate is a generalization of the arithmetic mean in an inner product space. A general notion of a mean of a probability distribution on a metric space was first due to [12]. The following can be viewed as a Fréchet mean with respect to the empirical distribution on $\{p_1,\ldots,p_N\}$ and the intrinsic distance $\mathrm{d}(\cdot,\cdot)$.

**Definition 5.** Let $D = \{p_1,\ldots,p_N\} \subset M$. The *intrinsic mean* of $D$ is

$$\mu(D) = \underset{x \in M}{\operatorname{argmin}} \frac{1}{N} \sum_{i=1}^{N} \mathrm{d}(x, p_i)^2.$$

The intrinsic mean is used as an offset in the following definition.

**Definition 6.** Given data $D = \{p_1,\ldots,p_N\} \subset M$ with intrinsic mean $\mu(D)$, the *intrinsic variance* of $D$ is

$$\sigma^2(D) = \frac{1}{N} \sum_{i=1}^{N} \mathrm{d}\{\mu(D), p_i\}^2.$$

*Principal geodesic analysis* (PGA), as introduced in [10], is generalization of principal component analysis (PCA) and is defined below.

**Definition 7.** Let $D = \{p_1,\ldots,p_N\} \subset M$ and $K = \dim(\mathrm{T}_{\mu(D)} M)$. PGA locates $\{v_1,\ldots,v_K\} \subset \mathrm{S}_{\mu(D)} M$ such that

$$v_1 = \underset{\|v\|=1}{\operatorname{argmin}} \frac{1}{N} \sum_{i=1}^{N} \mathrm{d}\{p_i, \pi_{H_1(v)}(p_i)\}^2 \text{ with } H_1(v) = \mathrm{Exp}_{\mu(D)}\{\mathrm{span}(v)\}$$

$$v_2 = \underset{\|v\|=1,\, v \in C_1}{\operatorname{argmin}} \frac{1}{N} \sum_{i=1}^{N} \mathrm{d}\{p_i, \pi_{H_2(v)}(p_i)\}^2$$

$$\vdots \qquad \vdots$$

$$v_K = \underset{\|v\|=1,\, v \in C_{K-1}}{\operatorname{argmin}} \frac{1}{N} \sum_{i=1}^{N} \mathrm{d}\{p_i, \pi_{H_K(v)}(p_i)\}^2$$

where $V_j = \{v_1,\ldots,v_j\}$, $C_j = \mathrm{span}(V_j)^\perp$ and $H_j(v) = \mathrm{Exp}_{\mu(D)}[\mathrm{span}(\{V_{j-1} \cup v\})]$ for $j = 2,\ldots,K$.

Objective functions in PCA are sum of squared distances of data to their orthogonal projections to linear subspaces. The symmetric, linear operator defined below is the gradient of the objective function in PCA.

**Definition 8.** Given $\{q_1,\ldots,q_N\} \in \mathrm{T}_\mu M$ define

$$\mathrm{L} : \mathrm{T}_u M \mapsto \mathrm{T}_u M, \quad \mathrm{L}(v) = \frac{2}{N} \sum_{i=1}^{N} \langle q_i, v \rangle_\mu q_i.$$

Throughout we assume that $\{q_1,\ldots,q_N\}$ is distributed so that the eigenvalues of L are distinct. Thus we can assume that the eigenvectors of L form an orthonormal set. We denote the eigenvectors of L by $u_1,\ldots,u_K$ with corresponding eigenvalues $\beta_1,\ldots,\beta_K$ given in descending order by magnitude.

## 3. Expansion of PGA directions

Let $S_r^n$, $P(n)$ and $SO(n)$ denote the $n$-sphere of radius $r$, the space of positive definite matrices and the special orthogonal group formulated as Riemannian manifolds as in Sections 4.1, 5.1, and 6.1, respectively. Using the above three spaces as working examples, in the following Proposition we explore the effects of scaling the Riemannian logs of the data in the tangent space. *Assume $M$ in the proposition in this section is one of these spaces.*

We let $\epsilon > 0$ be the *scaling parameter*. The dispersion of data from the intrinsic mean, $\mu(D)$, depends on the norm of each data point's Riemannian log in the tangent space at $\mu(D)$. As the data becomes more greatly dispersed in this manner the effects of curvature away from the tangent space should become more significant to the solutions to PGA and its components. By simultaneously scaling the Riemannian logs of all the data by $\epsilon$ and by obtaining Taylor expansions in this parameter we can discern and inspect this effect. In practice, such as in the simulations in Section 4.3 or in the improvement over the linear difference indicators in Section 7, we can take $\epsilon = 1$ with the norms of the Riemannian logs of the data determining the data's scale.

**Proposition.** Let $\mu \in M$, $\{q_1, \ldots, q_N\} \in T_\mu M$ and $u_1, \ldots, u_K$ be the eigenvectors of covariance operator L as in Definition 8. Also letting, for all $i \in \{1, \ldots, N\}$,
$$p_{i,\epsilon} = \operatorname{Exp}_\mu(\epsilon q_i)$$
let $D_\epsilon = \{p_{j,\epsilon}\}_j$. Assume $\mu(D_\epsilon) = \mu$ and that
$$V_{K,\epsilon} = \{v_1(\epsilon), \ldots, v_K(\epsilon)\}$$
is the set of PGA directions of $D_\epsilon$ for all $\epsilon \neq 0$.

Further, let $f_k(v, \epsilon)$ be the objective function for $v_k(\epsilon)$ in Definition 7 for $k = 1, \ldots, K$, i.e.,
$$f_k(v, \epsilon) = \frac{1}{N} \sum_{i=1}^{N} \mathrm{d}\{p_{i,\epsilon}, \pi_{H_k(v)}(p_{i,\epsilon})\}^2.$$

Then provided projection as in Definition 4, is unique, $f_k(v, \epsilon)$ is even in $\epsilon$ and we expand
$$f_k(v, \epsilon) = f_{k,2}(v)\epsilon^2 + f_{k,4}(v)\epsilon^4 + O(\epsilon^6). \tag{2}$$

Let $g_{k,4}(v)$ be as $f_{k,4}(v)$ with $u_1, \ldots, u_{k-1}$ in place of previous PGA directions, $v_1(\epsilon), \ldots, v_{k-1}(\epsilon)$. Also, for $j > k$ let
$$\alpha_k = \nabla_{u_k} g_{k,4} \quad \text{and} \quad \alpha_{k,j} = \langle \alpha_k, u_j \rangle = d_{u_k} g_{k,4}(u_j). \tag{3}$$

If $C$ is the $K \times K$ skew-symmetric matrix $C = (c_{k,j})$ where, for $j > k$,
$$c_{k,j} = \frac{\alpha_{k,j}}{\beta_j - \beta_k}$$

expanding $v_k(\epsilon) = v_{k,0} + v_{k,2}\epsilon^2 + O(\epsilon^4)$ then yields, for all $k \in \{1,\ldots,K\}$,

$$v_k(\epsilon) = v_{k,0} + v_{k,2}\epsilon^2 + O(\epsilon^4) = u_k + \left(\sum_{j=1}^{K} c_{k,j} u_j\right)\epsilon^2 + O(\epsilon^4).$$

PROOF. The proof is by induction. The base case can be shown in a similar manner as the induction step. Thus we let $k > 1$, assume the proposition holds for the first $k-1$ PGA directions $V_{k-1,\epsilon} = \{v_1(\epsilon),\ldots,v_{k-1}(\epsilon)\}$ and then show it holds for $v_k(\epsilon)$.

As in Sections 4.1, 5.1, and 6.1, $M$ is a symmetric space. It thus follows that $f_k(v,\epsilon)$ is even in $\epsilon$ as we assume projection is unique and the mapping

$$\iota : M \circlearrowleft, \ \iota(p) = \mathrm{Exp}_\mu\{-\mathrm{Log}_\mu(p)\} \quad \text{for } p \in M \tag{4}$$

is an isometry.

In Riemannian manifold $M$ intrinsic distances between points local to $\mu$ are Euclidean distances between Riemannian logs of these points in $\mathrm{T}_\mu M$ and thus we have

$$f_{k,2}(v) = \frac{1}{N}\sum_{i=1}^{N}\left(\langle q_i, q_i\rangle - \sum_{j=1}^{k-1}\langle q_i, u_j\rangle^2 - \langle q_i, v\rangle^2\right). \tag{5}$$

as the leading term of $f_k(v,\epsilon)$. Computing a gradient

$$\nabla_v f_{k,2} = -\frac{2}{N}\sum_{i=1}^{N}\langle q_i, v\rangle q_i = -\mathrm{L}(v)$$

so that

$$\nabla_v f_k(\epsilon) = -\mathrm{L}(v)\epsilon^2 + \nabla_v f_{k,4}\epsilon^4 + O(\epsilon^6) \tag{6}$$

Using Lagrange multipliers $(-\lambda_1,\ldots,-\lambda_{k-1},-(1/2)\lambda_k)$ we have

$$\nabla_{v_k(\epsilon)} f_k(\epsilon) = -\lambda_1 v_1(\epsilon) - \cdots - \lambda_k v_k(\epsilon) \tag{7}$$

with constraints

$$\langle v_1(\epsilon), v_k(\epsilon)\rangle = \cdots = \langle v_{k-1}(\epsilon), v_k(\epsilon)\rangle = 0, \langle v_k(\epsilon), v_k(\epsilon)\rangle = 1.$$

Expand $v_k(\epsilon)$ in $\epsilon$, viz.

$$v_k(\epsilon) = v_{k,0} + v_{k,2}\epsilon^2 + O(\epsilon^4). \tag{8}$$

Substituting this expansion and the expansions of $\{v_1(\epsilon),\ldots,v_{k-1}(\epsilon)\}$ in the constraints and equating coefficients in orders of $\epsilon$ gives

$$\begin{aligned}&\langle v_{k,0}, v_{j,0}\rangle = 0, \langle v_{k,0}, v_{j,2}\rangle = -\langle v_{k,2}, v_{j,0}\rangle \ \text{for } j = 1,\ldots,k-1,\\ &\langle v_{k,2}, v_{k,0}\rangle = 0 \ \text{and} \ \langle v_{k,0}, v_{k,0}\rangle = 1.\end{aligned} \tag{9}$$

For each $j \in \{1, \ldots, k\}$, expand the Lagrange multiplier

$$\lambda_j = \lambda_{j,0} + \lambda_{j,2}\epsilon^2 + \lambda_{j,4}\epsilon^4 + O(\epsilon^6). \tag{10}$$

In computations in Sections 4.2, 5.3, and 6.2 we have

$$\nabla_{v_k(\epsilon)} f_{k,4} = \nabla_{v_{k,0}} g_{k,4} + O(\epsilon^2).$$

Using this and substituting expansions (8), (10) and of $\{v_1(\epsilon), \ldots, v_{k-1}(\epsilon)\}$ in (7), using (6) and equating coefficients in orders of $\epsilon$ gives

$$\lambda_{1,0}v_{1,0} + \cdots + \lambda_{k,0}v_{k,0} = 0 \quad (\Rightarrow \lambda_{1,0} = \cdots = \lambda_{k,0} = 0)$$
$$\lambda_{1,2}v_{1,0} + \cdots + \lambda_{k,2}v_{k,0} = \mathrm{L}(v_{k,0}) \tag{11}$$
$$\sum_{j=1}^{k}(\lambda_{j,2}v_{j,2} + \lambda_{j,4}v_{j,0}) = -\nabla_{v_{k,0}}g_{k,4} + \mathrm{L}(v_{k,2}). \tag{12}$$

As L is a symmetric linear operator, (11) and (9) give

$$\lambda_{j,2} = \langle \mathrm{L}(v_{k,0}), v_{j,0} \rangle = \langle v_{k,0}, \mathrm{L}(v_{j,0}) \rangle = \beta_j \langle v_{k,0}, v_{j,0} \rangle = 0 \tag{13}$$

for each $j \in \{1, \ldots, k-1\}$.

Thus by (11), $\mathrm{L}(v_{k,0}) = \lambda_{k,2}v_{k,0}$ and $v_{k,0}$ is a normalized eigenvector of L with eigenvalue

$$\lambda_{k,2} = \frac{2}{N} \sum_{i=1}^{N} \langle v_{k,0}, q_i \rangle^2.$$

Further, letting $U_{k-1} = \{u_0, \ldots, u_{k-1}\}$, as

$$\forall_{\epsilon \neq 0} \quad \forall_{v \in SV_{k-1,\epsilon}^\perp} \quad f_k\{v_k(\epsilon), \epsilon\} \leq f_k(v, \epsilon)$$

using (5)

$$\forall_{v \in SU_{k-1}^\perp} \quad \frac{1}{N} \sum_{i=1}^{N} \langle q_i, v_{k,0} \rangle^2 \geq \frac{1}{N} \sum_{i=1}^{N} \langle q_i, v \rangle^2.$$

Thus $v_{k,0}$ is the dominant normalized eigenvector of L so that $v_{k,0} = u_k$ and $\beta_k = \lambda_{k,2}$. Then with $\alpha_k$ as in (3), by above, (12), and (13) we have

$$(\mathrm{L} - \beta_k)v_{k,2} = \sum_{j=1}^{k} \lambda_{j,4} u_j + \alpha_k. \tag{14}$$

Consider the orthonormal expansion of $v_{k,2}$, viz.

$$v_{k,2} = \sum_{j=1, j \neq k}^{K} c_{k,j} u_j. \tag{15}$$

For $j < k$, using (9) and the form of the expansions of $\{v_1(\epsilon), \ldots, v_{k-1}(\epsilon)\}$

$$c_{k,j} = \langle v_{k,2}, u_j \rangle = -\langle u_k, v_{j,2} \rangle = -c_{j,k}.$$

Also, substituting (15) into (14) gives

$$\sum_{j=1, j \neq k}^{K} c_{k,j}(\beta_j - \beta_k)u_j = \sum_{j=1}^{k} \lambda_{j,4} u_j + \alpha_k$$

so that for $j > k$

$$c_{k,j} = \langle \alpha_k, u_j \rangle /(\beta_j - \beta_k) = \alpha_{k,j}/(\beta_j - \beta_k).$$

Thus $v_k(\epsilon)$ has the form given in the proposition and the proposition holds.

**Remark 3.1.** In $v_k(\epsilon), v_{k,0} = u_k$ minimizes $f_{k,2}(v)$ subject to $\langle v, v \rangle = 1$. Then $v_{k,2}\epsilon^2$ is the first term which accounts for $f_{k,4}(v)\epsilon^4$ in the objective function.

For $j > k$ the numerator of $c_{k,j}$, $\alpha_{k,j}$, reflects the sensitivity of $f_{k,4}(v)$ evaluated locally, that is, $g_{k,4}(v)$, to a change in direction from $u_k$ towards $u_j$ with a greater magnitude giving a greater "benefit" in minimizing $f_{k,4}(v)$ in the objective function. The magnitude of the denominator, which is the difference of the shares of the data's variability in the tangent space accounted for by $u_j$ and $u_k$, respectively, reflects the "cost," with respect to the minimization of $f_{k,2}(v)$, of this change in direction. Further, for $j < k$ with $v_k(\epsilon)$ minimizing the sum of square residuals after $v_1(\epsilon), \ldots, v_{k-1}$, $\alpha_{k,j}$ accounts for the sensitivity of $g_{j,4}(v)$ to a change from $u_j$ in the direction of $u_k$ but with an opposite sign than $\alpha_{j,k}$ making $C$ skew-symmetric.

Unlike the effect of the scaling parameter, $\epsilon$, which is made explicit with the PGA expansion, the role of curvature on the PGA directions is more subtle. In $f_k(v, \epsilon), f_{k,4}(v)$ is the first coefficient present due to non-linearity and thus generally greater curvature will create greater $\alpha_{k,j}'s$. Greater scale of the data as measured by the norms of the Riemannian logs of the data and by $\epsilon$ augments this effect with $v_k(\epsilon) = v_{k,0} + v_{k,2}\epsilon^2 + O(\epsilon^4)$. Although the direct characterization of dependence on curvature for general symmetric spaces is not available, for all of the examples we considered, as in (18), (19), (26), (29), curvature appears explicitly in this manner.

## 4. PGA in $S_r^n$

We first examine the role of scale in PGA in the $n$-spheres. We denote the $n$-spheres of radius $r$ by $S_r^n$. Spherical data occurs in directional statistics, in preshapes in shape analysis, in text mining, in cluster analysis and others as in [1], [14], [17], and [23]. We obtain the expansions of PGA directions in $S_r^n$ according to the proposition in Section 3, test these expansions and then apply them to simulated data to show improved approximations of PGA directions.

*4.1. $S_r^n$ as a Riemannian manifold*

We have the identification

$$\mathrm{T}_p S_r^n \equiv \{v \in \mathbb{R}^{n+1}; \langle v, p \rangle = 0\}.$$

On $S_r^n$ we use the Riemannian metric induced by the embedding $S_r^n \hookrightarrow \mathbb{R}^{n+1}$, i.e., for any $p \in S_r^n$ and $X, Y \in \mathrm{T}_p S_r^n$, $\langle X, Y \rangle$ is the dot product. The geodesics in $S_r^n$ are great circles and the *Riemannian exponential map in $S_r^n$* is directly computed as below.

$$\mathrm{Exp}_p(X) = \cos\left(\frac{\|X\|}{r}\right) p + r \sin\left(\frac{\|X\|}{r}\right) \frac{X}{\|X\|}$$

for $X \in \{Y \in \mathrm{T}_p S(n); \|Y\| < \pi r\}$. Then

$$\mathrm{Log}_p(q) = \begin{cases} 0, & \text{for } \langle p, q \rangle = r^2, \\ r \arccos\left(\langle p, q \rangle / r^2\right) \frac{q - \langle p, q \rangle p / r^2}{\|q - \langle p, q \rangle p / r^2\|} & \text{for } |\langle p, q \rangle| < r^2. \end{cases} \quad (16)$$

Given $p, q \in S_r^n$

$$\mathrm{d}(p, q) = \|\mathrm{Log}_p(q)\| = r \arccos\left(\langle p, q \rangle / r^2\right),$$

that is, the distance between $p$ and $q$ is the ordinary spherical distance.

As in [7], $S_r^n$ is a symmetric space with the symmetry at any point $p \in S_r^n$ provided by reflection over the line containing $p$ in $\mathbb{R}^{n+1}$. Also, we have $K(\sigma_{v,q}) = 1/r^2$ for any plane $\sigma_{v,q} \subset \mathrm{T}_p S_r^n$.

*4.2. Expansion of PGA directions in $S_r^n$*

As in [17] the projection operator in $S_r^n$ has closed-form. Given $p, \mu \in S_r^n$ and an orthonormal set $V_k = \{v_1, \ldots, v_k\} \subset \mathrm{T}_\mu S_r^n$, set $v_0 = \mu/r$. With $p \notin \mathrm{span}(\mu \cup V_k)^\perp$ so that projection is unique then

$$\pi_{H(V_k)}(p) = rw/\|w\|, \quad (17)$$

where $w = \langle v_0, p \rangle v_0 + \cdots + \langle v_k, p \rangle v_k$. That is, projection of $p$ is first done to $\mathrm{span}(\mu \cup V_k)$ in $\mathbb{R}^{n+1}$ and the result scaled to obtain projection to the hypersphere $H(V_k)$.

Let $q \in \mathrm{T}_\mu S_r^n$, $\epsilon \neq 0$, $p_\epsilon = \mathrm{Exp}_\mu(\epsilon q)$ and, for all $j \in \{1, \ldots, k\}$,

$$t_j(\epsilon) = \left\langle \mathrm{Log}_\mu\{\pi_{H(V_k)}(p_\epsilon)\}, v_j \right\rangle,$$

so that

$$\pi_{H(V_k)}(p_\epsilon) = \mathrm{Exp}_\mu\left\{\sum_{j=1}^k t_j(\epsilon) v_j\right\}.$$

By using (16) and (17), taking an inner product and computing Taylor expansions we obtain the following *expansion of projection coefficients in $S_r^n$*:

$$\begin{aligned}
t_m(\epsilon) &= \langle q, v_m \rangle \epsilon + \frac{\langle q, v_m \rangle}{3r^2}\left(\langle q, q \rangle^2 - \sum_{j=1}^k \langle q, v_j \rangle^2\right) \epsilon^3 + O(\epsilon^5) \\
&= \cos\theta_m \|q\| \epsilon + \frac{\cos\theta_m}{3r^2}\left(1 - \sum_{j=1}^k \cos^2\theta_j\right) \|q\|^3 \epsilon^3 + O(\epsilon^5)
\end{aligned} \quad (18)$$

for each $m \in \{1, \ldots, k\}$, where $\theta_m$ the angle formed by $q$ and $v_m$.

As in the proposition in Section 3 assume $\mu \in S_r^n$, $q_i \in T_\mu S_r^n$ and $p_{i,\epsilon} = \text{Exp}_\mu(\epsilon q_i)$ for each $i \in \{1, \ldots, N\}$ with $V_{k-1,\epsilon} = \{v_1(\epsilon), \ldots, v_{k-1}(\epsilon)\}$ the first $k-1$ PGA directions. Using the proposition we find $v_k(\epsilon)$. Let $v \in S_\mu S_r^n$, $V_{k,\epsilon} = \{V_{k-1,\epsilon} \cup v\}$ and

$$t_{i,j} = \langle \text{Log}_\mu\{\pi_{H(V_{k,\epsilon})}(p_{i,\epsilon})\}, v_j \rangle \quad \text{and} \quad t_{i,k} = \langle \text{Log}_\mu\{\pi_{H(V_{k,\epsilon})}(p_{i,\epsilon})\}, v \rangle$$

for all $i \in \{1, \ldots, N\}$ and $j \in \{1, \ldots, k-1\}$.

Our objective function is

$$f_k(v, \epsilon) = \frac{1}{N} \sum_{i=1}^N r^2 \text{acos}^2\left(\langle \text{Exp}_\mu(t_{i,1}v_1 + \cdots + t_{i,k}v), \text{Exp}_\mu(\epsilon q_i)\rangle / r^2\right)$$

$$= \frac{1}{N} \sum_{i=1}^N r^2 \text{acos}^2 \Bigg\{ \cos\left(\sqrt{\sum_{j=1}^k t_{i,j}^2}/r\right) \cos\left(\frac{\epsilon \|q_i\|}{r}\right) +$$

$$\sin\left(\sqrt{\sum_{j=1}^k t_{i,j}^2}/r\right) \sin\left(\frac{\epsilon \|q_i\|}{r}\right) \frac{\sum_{j=1}^{k-1} \langle t_{i,j}v_j, q_i\rangle + \langle t_{i,k}v, q_i\rangle}{\|q_i\| \sqrt{\sum_{j=1}^k t_{i,j}^2}} \Bigg\}.$$

Then by taking Taylor expansions in $\epsilon$ and with $f_{k,4}(v)$ as in (2) in the proposition in Section 3 we obtain

$$f_{k,4}(v) = \frac{1}{3Nr^2} \sum_{i=1}^N \Bigg\{ \left(\sum_{j=1}^{k-1} \langle q_i, v_j\rangle^2 + \langle q_i, v\rangle^2\right) \times$$

$$\left(\sum_{j=1}^{k-1} \langle q_i, v_j\rangle^2 + \langle q_i, v\rangle^2 - \langle q_i, q_i\rangle\right) \Bigg\}.$$

With $\alpha_{k,m}$ as in (3), taking a gradient and evaluating gives the following *expansion of $v_k'$s in $S_r^n$*.

$$\alpha_{k,m} = \langle \nabla_{u_k} g_{k,4}, u_m \rangle$$

$$= \frac{2}{3Nr^2} \sum_{i=1}^N \left(\sum_{j=1}^k 2\langle q_i, u_j\rangle^2 - \langle q_i, q_i\rangle\right) \langle q_i, u_k\rangle \langle q_i, u_m\rangle \quad (19)$$

$$= \frac{2}{3Nr^2} \sum_{i=1}^N \|q_i\|^4 \left(\sum_{j=1}^k 2\cos^2 \theta_{i,j} - 1\right) \cos\theta_{i,k} \cos\theta_{i,m},$$

where $\theta_{i,a}$ is the angle formed by $q_i$ and $u_a$ for all $i, a$ and which is used in the proposition to obtain, for each $k \in \{1, \ldots, n\}$, the expansion

$$v_k(\epsilon) = v_{k,0} + v_{k,2}\epsilon^2 + O(\epsilon^4).$$

Let $S^n$ be the unit sphere $S_1^n$. In Figure 2 the expansions of PGA directions obtained above are tested in $S^{10}$. With $\mu \in S^{10}$, 50 tangent vectors are sampled from $T_\mu S^{10}$ with the entries of $\{q_i\}$ having normal distributions and variances varying by entry so that principal geodesic directions can be identified. We then take $D_\epsilon = \{p_{i,\epsilon}\}_i = \{\mathrm{Exp}_\mu(\epsilon q_i)\}_i$. PGA directions are located in MATLAB by using fixed-point algorithms such as those in [17] used to compute *principal component geodesics*. Ln-ln plots are shown for $v_1(\epsilon), v_2(\epsilon), v_4(\epsilon)$ and $v_9(\epsilon)$. Similar plots were obtained for expansions of the other PGA directions.

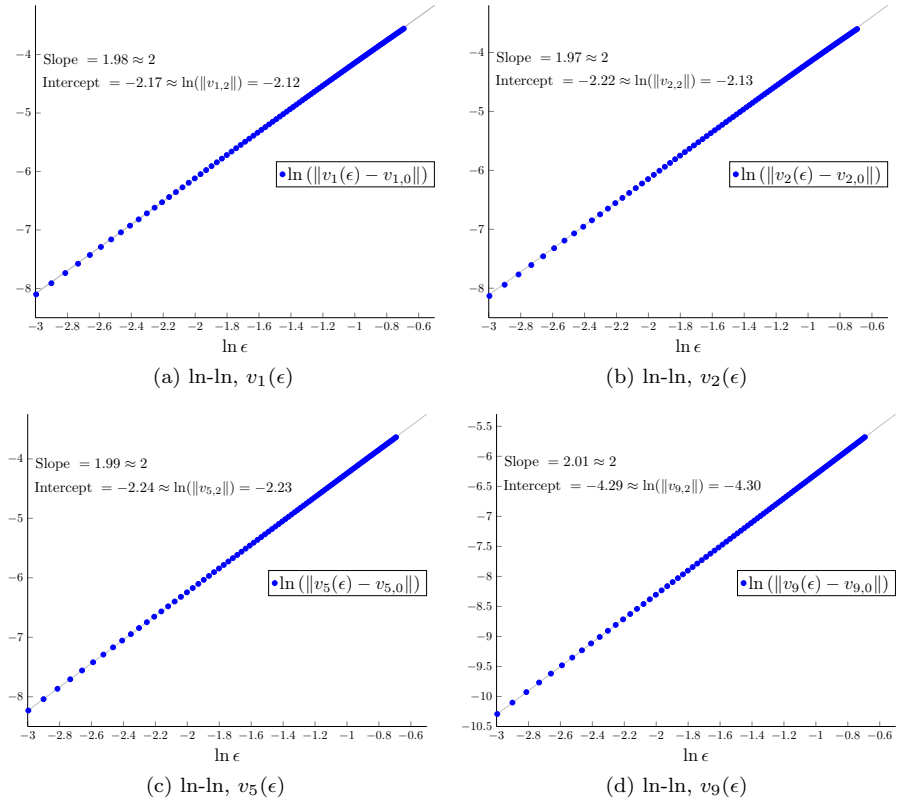

Figure 2: Tests of PGA expansions in $S^{15}$

*4.3. Approximations of PGA on Simulated Data in $S^n$*

In this section we sample data from a generalization of the log-normal distribution in $S^{15}$. Specifically we sample from $X$ where $X \sim \mathrm{Exp}_\mu(\mathcal{N}(0, \kappa\Sigma))$ with $\mu \in S^{15}$, $\kappa > 0$ and $\Sigma$ a fixed positive definite matrix. $\Sigma$ has distinct eigenvalues with the largest eigenvalue having 20 times the magnitude of the smallest so that our distribution on $S^{15}$ is anisotropic and the population PGA directions exist. We compare PGA directions to approximations while varying $\kappa$ which positively correlates with the expected dispersion of the data from its mean.

In Table 1 for each value of $\kappa$ 100 data points are sampled in 5 runs each. For each run all 14 PGA directions are computed. Also, their leading- and next-order approximations in the scale of the data are located, that is, $v_{k,0}$ (using

PCA in the tangent space) and $v_{k,0} + v_{k,2}$, respectively, with $\epsilon = 1$ as the norms of the Riemannian logs of the data scale the approximations. Across all runs and for each value of $\kappa$ the means of the angles in radians the approximations make (m.est. $\theta_0$ and m.est. $\theta_2$) with the exact PGA directions and the mean of the norms of the Riemannian logs of the data (m.scale) are computed.

Also, as proposed in [29], an iterative algorithm to locate PGA direction $v_k$ can be initialized by doing PCA with projections of the Riemannian logs of the data into $\text{span}(V_{k-1}^\perp)$. We compare this method with initialization by projection of the next-order approximations into $\text{span}(V_{k-1}^\perp)$ by computing the means of the angles these initializations make (m.init.$\theta_0$ and m.init.$\theta_2$, respectively) with the $v'_j s$ across the 5 runs for each value of $\kappa$.

| $\kappa$ | 1 | .850 | .700 | .550 | .400 | .250 | .100 | .050 |
|---|---|---|---|---|---|---|---|---|
| m.scale | 1.4684 | 1.3361 | 1.2278 | 1.1867 | 0.9784 | 0.7472 | 0.4843 | 0.3399 |
| m.est.$\theta_0$ | 0.3400 | 0.2558 | 0.2327 | 0.1659 | 0.1442 | 0.0921 | 0.0576 | 0.0190 |
| m.est.$\theta_2$ | 0.2654 | 0.2323 | 0.1595 | 0.0959 | 0.0541 | 0.0371 | 0.0215 | 0.0017 |
| m.init.$\theta_0$ | 0.2329 | 0.1717 | 0.1573 | 0.1146 | 0.0972 | 0.0613 | 0.0347 | 0.0123 |
| m.init.$\theta_2$ | 0.1882 | 0.1528 | 0.1014 | 0.0636 | 0.0347 | 0.0225 | 0.0120 | 0.0010 |

m.scale = mean of norms of Riemannian logs
m.est.$\theta_0$, est.$\theta_2$ = mean angles of leading and next order estimates w/$v'_j s$
m.init.$\theta_0$, init.$\theta_2$ = mean angles of leading and next order initializations w/$v'_j s$

Table 1: Estimates of PGA in $S^{15}$ from Log-normal samples

At $\kappa = 1$ the data is nearly uniformly distributed which is reflected in m.scale = 1.4684 nearly $\pi/2$. Even at $\kappa = 1$ the next-order estimates are an improvement over the leading-order estimates for this distribution. Note that the next-order estimates holding for data dispersed significantly from the the mean is also reflected in the plots with nearly correct intercepts at $\ln(\epsilon) = 0 \Rightarrow \epsilon = 1$ in Figure 2. As $\kappa$ decreases and the samples draw in towards their computed means both estimates improve with the next-order estimates improving more sharply as they should.

Also, projecting the next-order estimates into $\text{span}(V_{k-1}^\perp)$ provides an improvement here over doing PCA in $\text{span}(V_{k-1}^\perp)$ with m.init. $\theta_2$ < m.init. $\theta_0$ for all values of $\kappa$.

In Figure 3, $\kappa = .4$ and for each PGA direction we plot the values of est. $\theta_0$, est.$\theta_2$ and init. $\theta_0$, init.$\theta_2$ which are means across 5 runs. Similar plots were obtained for other values of $\kappa$.

## 5. PGA in $P(n)$

We denote the space of positive definite matrices by $P(n)$. The imaging technology diffusion tensor MRI as in [2] produces data known as diffusion tensors in $P(3)$. In [10] PGA in Definition 7 was proposed to allow the proper analysis of statistical variability of diffusion tensor data by formulating $P(3)$ as a manifold and generalizing PCA.

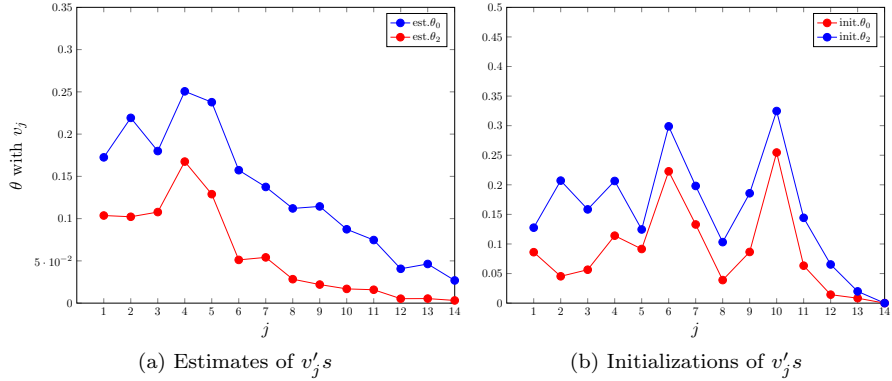

Figure 3: Estimates of PGA in $S^{15}$ with $\kappa = .4$

In this section using the proposition in Section 3 and computer algebra we obtain expansions of PGA directions in $P(n)$. We test the expansions and we take a closer look at the geometry of the first projection coefficient.

*5.1. $P(n)$ as a Riemannian manifold*

As in [22], $P(n)$ is an open set in the vector space of $n \times n$ symmetric matrices. Thus, we have the identification

$$\mathrm{T}_p P(n) \equiv n \times n \text{ symmetric matrices}.$$

Denote the general linear group by $GL(n)$. Consider the following *action on $P(n)$*, $\varphi$

$$\varphi : GL(n) \times P(n) \to P(n), \quad \varphi(g,p) = \varphi_g(p) = gpg^\top. \tag{20}$$

As in [3] this action is transitive. We have the following *Riemannian metric on $P(n)$* for which $\varphi_g$ is an isometry up to a positive scalar multiple.

$$\langle X, Y \rangle_p = (1/2)\mathrm{tr}\left(p^{-1} X p^{-1} Y\right)$$

for $p \in P(n), X, Y \in \mathrm{T}_p P(n)$ and where tr denotes the matrix trace.

As in [3] setting $M = P(n)$ and $\phi_p(q) = pq^{-1}p$ for $q \in P(n)$ in Definition 1 makes $P(n)$ a symmetric space.

For vector fields $x, y$ on $P(n)$, $[x,y] = xy - yx$ is the *commutator* of $x, y$. Also, as $P(n)$ is a symmetric space, by [25], we have

$$\mathrm{R}(x,y)z = [z, [x,y]] \tag{21}$$

As in [3], at $\mathrm{I} \in P(n)$, $\mathrm{Exp}_\mathrm{I}$ is the matrix exponential function and we have the following *Riemannian exponential map on $P(n)$*, for $X \in \mathrm{T}_\mathrm{I} P(n)$,

$$\mathrm{Exp}_\mathrm{I}(X) = \sum_{i=0}^{\infty} X^i/i! \tag{22}$$

and for any $p = gg^\top \in P(n)$, where $g \in GL(n)$ we have

$$\mathrm{Exp}_p(X) = \phi_g[\mathrm{Exp}_I\{d_p\phi_{g^{-1}}(X)\}] = g\,\mathrm{Exp}_I\{g^{-1}X(g^{-1})^\top\}g^\top$$

for $X \in \mathrm{T}_p P(n)$.

Also in [3], for all $p \in P(n)$, $\mathrm{Exp}_p$ is bijective and thus $\mathrm{Log}_p$ is well-defined.

*5.2. Computer Algebra*

Expansions in this section and in Sections 6.3 and 7 were obtained with the help of the Maxima computer algebra system. A matrix Taylor function and a function with the distributive and cyclic properties of the trace were coded to compute expansions involving $\mathrm{Log}_I$ and $\mathrm{Exp}_I$ in $\epsilon$. This code along with the data sets used in this paper are available at https://github.com/DMLazar/PGAScale.

*5.3. Expansion of PGA directions in $P(n)$*

By employing the transitive action by isometries in (20) PGA in $P(n)$ can be carried out in the tangent space at the identity. Let $v \in \mathrm{S}_I P(n), q \in \mathrm{T}_I P(n)$ and $p_\epsilon = \mathrm{Exp}_I(\epsilon q)$ for $\epsilon \neq 0$. To project $p_\epsilon$ to $H(v) = \mathrm{Exp}_I\{\mathrm{span}(v)\}$ find

$$t(\epsilon) = \underset{s \in \mathbb{R}}{\mathrm{argmin}}\ \mathrm{d}\{\mathrm{Exp}_I(sv), \mathrm{Exp}_I(\epsilon q)\}^2 \qquad (23)$$

Define $O_{\epsilon,s}(\ell)$ as

$$f(s,\epsilon) \text{ is } O_{\epsilon,s}(\ell) \quad \Leftrightarrow \quad f(s,\epsilon) \leq \sum_{k=0}^{\ell} A_k \epsilon^k s^{\ell-k}$$

for some $A_1, \ldots, A_\ell \in \mathbb{R}$. Then let

$$g(s,\epsilon) = \mathrm{Exp}_I(-sv/2)\mathrm{Exp}_I(\epsilon q)\mathrm{Exp}_I(-sv/2)$$

and setting $h(s,\epsilon)$ as the cost function in (23) and expanding

$$\begin{aligned}
h(s,\epsilon) &= (1/2)\mathrm{tr}[\mathrm{Log}_I\{g(s,\epsilon)\}\mathrm{Log}_I\{g(s,\epsilon)\}] \\
&= \frac{\epsilon^2 \mathrm{tr}\,(qq)}{2} + s^2 - \epsilon s\,\mathrm{tr}\,(qv) + \frac{\mathrm{tr}\{q^2 v^2 - (qv)^2\}\epsilon^2 s^2}{12} + O_{\epsilon,s}(6).
\end{aligned} \qquad (24)$$

Using (4), $t(\epsilon)$ is odd in $\epsilon$ and we expand

$$t(\epsilon) = t_1 \epsilon + t_3 \epsilon^3 + O(\epsilon^5)$$

for some $t_1, t_3 \in \mathbb{R}$. Solving for $t_1$ and $t_3$ in

$$\frac{\partial h_s(t(\epsilon),\epsilon)}{\partial s} = 0$$

gives

$$t(\epsilon) = \frac{1}{2}\mathrm{tr}\,(qv)\,\epsilon + \frac{1}{24}\mathrm{tr}\,(qv)\left[\mathrm{tr}\{(qv)^2 - q^2 v^2\}\right]\epsilon^3 + O(\epsilon^5). \qquad (25)$$

Using (1) in (25) we have the *expansion of a projection (to a geodesic) coefficient in $P(n)$* below

$$t(\epsilon) = \frac{1}{2}\text{tr}\,(qv)\,\epsilon + \frac{1}{24}\text{tr}\,(qv)\left[\text{tr}\{(qv)^2 - q^2v^2\}\right]\epsilon^3 + O(\epsilon^5)$$
$$= \cos\theta\,\|q\|\,\epsilon + \frac{1}{12}\cos\theta\sin^2\theta K(\sigma_{v,q})\,\|q\|^3\,\epsilon^3 + O(\epsilon^5), \quad (26)$$

where $\theta$ is the angle formed by $q$ and $v$.

In Figure 4 $v$ and $q$ are sampled from the uniform distribution on the unit sphere in $\text{T}_\text{I}P(n)$ and plots are generated in MATLAB to test the expansions.

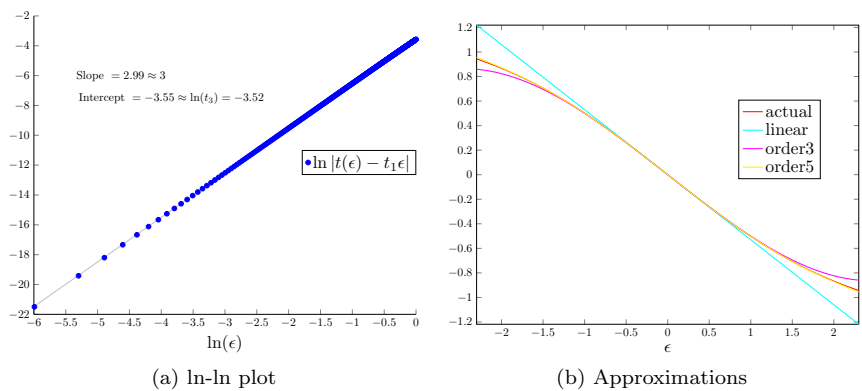

(a) ln-ln plot    (b) Approximations

Figure 4: Tests of expansion of $t(\epsilon)$

As in Figure 5, letting $\|q\| = 1$ so that $|\epsilon| = \text{d}(p_\epsilon, \text{I})$, as $\epsilon$ goes to zero the tangent vectors become more like their exponents and projection in the tangent space becomes more like projection in the manifold. The third-order term accounts for the difference in the approximating Euclidean triangle in the tangent space and the geodesic triangle in the manifold. As in [3] $P(n)$ is of non-positive sectional curvature. Thus $\cos\theta\sin^2\theta K(\sigma_{v,q})/12$ will be non-positive for acute angle $\theta$ with greater local curvature of $\text{Exp}_\mu\{\text{span}(v,q)\}$, as measured by the magnitude of $K(\sigma_{v,q})$, contributing to a less accurate approximation.

We apply the proposition to compute the expansions of the first PGA directions in $P(n)$. First we have

$$f_1(v,\epsilon) = \frac{1}{N}\sum_{i=1}^{N}\text{d}[\text{Exp}_\text{I}\{t_i(\epsilon,v)v\}, \text{Exp}_\text{I}(\epsilon q_i)]^2 \quad (27)$$

With $f_{1,4}(v)$ as in (2) in the proposition in Section 3, using (25) in (27) and expanding in $\epsilon$ gives

$$f_{1,4}(v) = \frac{1}{48N}\sum_{i=1}^{N}\text{tr}\,(q_iv)^2\,\{\text{tr}\,(q_i^2v^2) - \text{tr}\,(q_ivq_iv)\}. \quad (28)$$

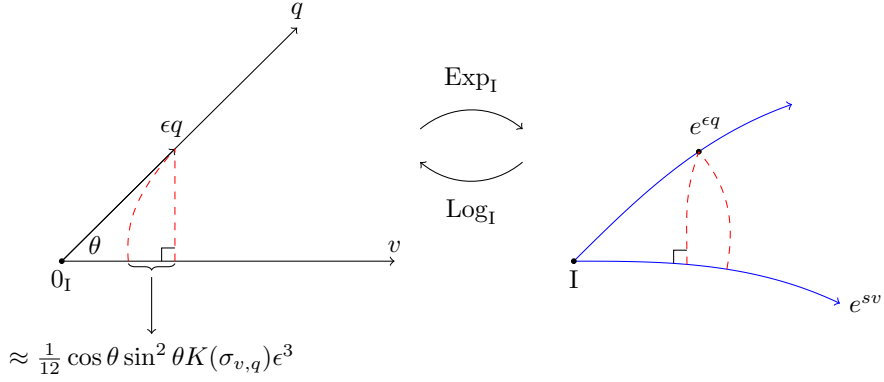

Figure 5: Approximation of projection coefficient

With $\alpha_{1,j}$ as in (3), taking a gradient and evaluating gives the following *expansion of $v_1(\epsilon)$ in $P(n)$*.

$$\begin{aligned}
\alpha_{1,j} &= \langle \nabla_{u_1} g_{1,4}, u_j \rangle = (1/2)\text{tr}\,(\alpha_1 u_j) \\
&= \frac{1}{48N} \sum_{i=1}^{N} \text{tr}\,(q_i u_1)^2 \,\text{tr}\,\left(q_i^2 u_1 u_j + q_i^2 u_j u_1 - 2q_i u_1 q_i u_j\right) \\
&\quad + 2\text{tr}\,(q_i u_1)\,\text{tr}\,(q_i u_j)\,\text{tr}\{q_i^2 u_1^2 - (q_i u_1)^2\} \\
&= -\frac{1}{6N} \sum_{i=1}^{N} \|q_i\|^4 \left\{ \cos^2 \theta_{i,1} R_I(u_1, \tilde{q}_i, \tilde{q}_i, u_j) \right. \\
&\quad \left. + \cos \theta_{i,1} \cos \theta_{i,j} R_I(u_1, \tilde{q}_i, \tilde{q}_i, u_1) \right\} \\
&= -\frac{1}{6N} \sum_{i=1}^{N} \|q_i\|^4 \left\{ \cos^2 \theta_{i,1} R_I(u_1, \tilde{q}_i, \tilde{q}_i, u_j) \right. \\
&\quad \left. + \cos \theta_{i,1} \cos \theta_{i,j} \sin^2 \theta_{i,1} K(\sigma_{q_i, u_1}) \right\}
\end{aligned} \qquad (29)$$

where $\theta_{i,m}$ is the angle formed by $q_i$ and $u_m$ and $\tilde{q}_i = q_i/\|q_i\|$ for all $i, m$ and which can be used in the proposition to obtain the expansion of $v_1(\epsilon)$.

PGA directions in addition to the first can be found by first solving for projection coefficients in systems of equations. As in (29) substituting the projection coefficients in the PGA objective function, taking a gradient and evaluating gives $\alpha_{k,j}$. At https://github.com/DMLazar/PGAScale the Maxima code for and the form of $v_2(\epsilon)$ can be found.

In Figure 6, the expansions of $v_1(\epsilon)$ and $v_2(\epsilon)$ in $P(n)$ are tested. 75 matrices $\{q_i\}_i$ are sampled from $T_I P(3)$ with entries distributed as in the test in Figure 2 and the data is set as $D_\epsilon = \{p_{i,\epsilon}\}_i = \{\text{Exp}_I(\epsilon q_i)\}_i$. The computations of the projection operator and principal geodesic directions are done using MATLAB minimization routines and user-supplied gradients as formulated in [29] with the derivative of the matrix exponential map provided by [24, Theorem 4.5].

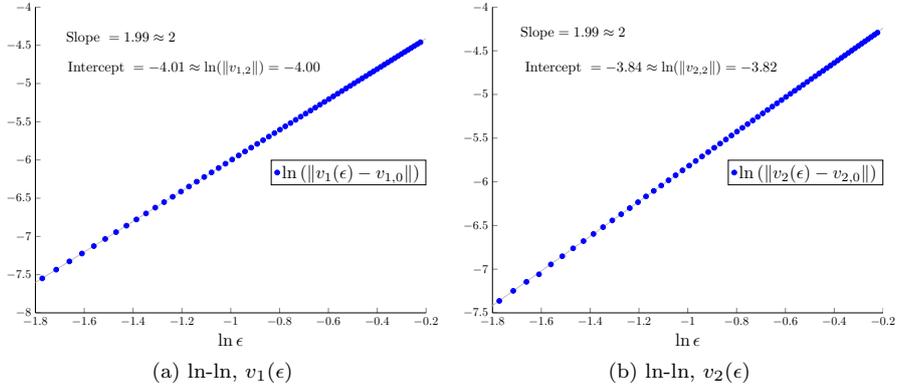

(a) ln-ln, $v_1(\epsilon)$

(b) ln-ln, $v_2(\epsilon)$

Figure 6: Tests of PGA expansions in $P(n)$

## 6. PGA in $SO(n)$

We denote the *special orthogonal group* by $SO(n)$. $SO(n)$ is the group of rigid rotations of $\mathbb{R}^n$. In particular it includes *the rotation group $SO(3)$* in which data naturally arises in robotics, computer vision and others (see [19], [21] and [30]). We obtain expansions in $SO(n)$ and apply them to the formulation of PGA in [11] for Lie groups. We use the identification of $SO(3)$ with $SU(2)$ as in [27] and fixed-point algorithms as in [17] to carry out PGA.

### 6.1. $SO(n)$ as a Riemannian manifold

As in [5] $SO(n)$ is a closed and bounded subset of $\mathbb{R}^{n^2}$ and is thus a compact set. Also, we have the identification

$$\mathrm{T}_I SO(n) \equiv \mathfrak{so}(\mathfrak{n}) = \{n \times n \text{ skew-symmetric matrices}\}.$$

$SO(n)$ is a matrix Lie group and we have the following transitive *action on $SO(n)$*, $\varphi$ by left multiplication

$$\varphi : SO(n) \times SO(n) \to SO(n),$$
$$\varphi(g, p) = \varphi_g(p) = gp.$$

We then have the Riemannian metric for which $\varphi_g$ is an isometry up to a positive scalar multiple.

$$\langle X, Y \rangle_p = -(1/2)\mathrm{tr}\left(p^{-1} X p^{-1} Y\right) \qquad (30)$$

for $p \in SO(n), X, Y \in \mathrm{T}_p SO(n)$.

As in $P(n)$ setting $M = SO(n)$ and $\phi_p(q) = pq^{-1}p$ for $q \in SO(n)$ in Definition (1) makes $SO(n)$ a symmetric space. Further, as $SO(n)$ is a symmetric space we have (21) as in $P(n)$.

As in [8], at I ∈ $SO(n)$, $\text{Exp}_I$ is the matrix exponential function and we have the *Riemannian exponential map on $SO(n)$*, defined, for $X \in \{Y \in T_I SO(n)$ such that $\|Y\| < \pi\}$ by (22) and for any $p \in SO(n)$ we have

$$\text{Exp}_p(X) = \phi_p \left[ \text{Exp}_I\{d_p \phi_{p^{-1}}(X)\} \right] = p \text{Exp}_I(p^{-1}X)$$

for $X \in \{Y \in T_p SO(n)$ such that $\|Y\| < \pi\}$.

Also, using the results in [13], for any $p$, $\text{Exp}_p$ is a diffeomorphism on $\{X \in T_p SO(n); \|X\| < \pi\}$.

### 6.2. Expansion in $SO(n)$

Both $P(n)$ and $SO(n)$ are symmetric spaces that have the matrix exponential as the Riemannian exponential map at I. At I, the inner products of $P(n)$ and $SO(n)$ are scalar multiples of the trace of matrix products of tangent vectors. Also, the action in (20), with $SO(n)$ replacing $GL(n)$ and $P(n)$, is an action by isometries which is transitive as any $p \in SO(n)$ can be decomposed as $p = gg$ where $g = \text{Exp}_I\{(1/2)\text{Log}_I(p)\}$. Thus, the expansions in $\epsilon$ in $SO(n)$ are the expansions in Section 5.3 for $P(n)$ with the inner product of $SO(n)$ replacing the inner product of $P(n)$.

### 6.3. Lie group PGA

By use of expansions in scaling parameter $\epsilon$, in the case of $SO(n)$, we take a look at the first iteration of PGA in [11]. In a Lie group such as $SO(n)$, PGA was formulated in [11] as recursive variance removal through left multiplication as below.

**Definition 9** (PGA alternative). Given Lie group $M$ and $D = \{p_1, \ldots, p_N\} \subset M$ with $\mu(D) = \mu$, PGA directions $\{v_1, \ldots, v_K\} \subset S_{\mu(D)} M$ are given

Set $k = 1$.

I. Find $v_k = \underset{\|v\|=1}{\text{argmin}} \sum d\{\pi_{H(v)}(p_i), p_i\}^2$ with $H(v) = \text{Exp}_\mu\{\text{span}(v)\}$

II. Set $D' = \{g_1^{-1} p_1, \ldots, g_N^{-1} p_N\}$ where $g_i = \pi_{H(v_k)}(p_i)$ for all $k$

III. If $k < K$ set $k = k+1, D = D'$ and return to I, else stop.

In $SO(n)$ we can decompose any projection $g_i = \text{Exp}_\mu(t_i v_k)$ above as

$$g_i = \text{Exp}_\mu(a t_i v_k) \text{Exp}_\mu(b t_i v_k),$$

where $a + b = 1$. Then $\gamma_{a,b} : p \to \text{Exp}_\mu\{-a(t_i v_k)\} p \text{Exp}_\mu\{-b(t_i v_k)\}$ is an isometry as for all $X \in T_\mu SO(n)$,

$$d_\mu \gamma_{a,b}(X) = \text{Exp}_\mu\{-a(t_i v_k)\} X \text{Exp}_\mu\{-b(t_i v_k)\}$$

and through direct computation and using (30) we have, for all $X, Y \in T_\mu SO(n)$,

$$\langle d_\mu \gamma_{a,b}(X), d_\mu \gamma_{a,b}(Y) \rangle_{\gamma_{a,b}(\mu)} = \langle X, Y \rangle_\mu.$$

Of these isometries we should choose the one that "minimizes energy" in moving parts of data along the geodesic from $g_i$ to $\mu$. Assuming $D$ has been demeaned so that $\mu = \mathrm{I}$, letting $p_i = \mathrm{Exp}_\mathrm{I}(\epsilon q_i)$ and expanding

$$\mathrm{Log}_\mathrm{I}[\gamma_{a,b}\{\mathrm{Exp}_\mathrm{I}(\epsilon q_i)\}]$$
$$= \{q_i + (1/2)\mathrm{tr}\,(q_i v_k)\,v_k\}\epsilon - (1/4)\mathrm{tr}\,(q_i v_k)\,(b-a)(v_k q_i - q_i v_k)\epsilon^2 + O(\epsilon^3)$$
$$= \|q_i\|\,(\tilde{q}_i - \cos\theta_i v_k)\epsilon + \|q_i\|^2\,(1/2)\cos\theta_i(b-a)[v_k, \tilde{q}_i]\epsilon^2 + O(\epsilon^3)$$

where $\theta_i$ is the angle formed by $q_i$ and $v_k$. Thus for $a \neq b$ we have additional movement in the orthogonal complement of explanatory direction $v_k$. Note that accordingly, making the identification of $SO(3)$ with $SU(2)$, it is straightforward to show that with $a = b$, $\gamma_{a,b}$ corresponds to a simple plane rotation.

Also to consider in this method of PGA is the displacement of the mean as a result of curvature after removing variability in an explanatory direction. We quantify this effect in $SO(n)$ by letting $D_\epsilon = \{p_{1,\epsilon}, \ldots, p_{N,\epsilon}\} \subset SO(n)$ be such that $\mu(D_\epsilon) = \mathrm{I}$ for all $\epsilon, v_k \in \mathrm{S}_\mathrm{I}SO(n)$ and

Case 1. $D'_\epsilon = \{g_1^{-1/2} p_{1,\epsilon} g_1^{-1/2}, \ldots, g_N^{-1/2} p_{N,\epsilon} g_N^{-1/2}\}$ using $a = b = 1/2$ and

Case 2. $D'_\epsilon = \{g_1^{-1} p_{1,\epsilon}, \ldots, g_N^{-1} p_{N,\epsilon}\}$ using $a = 1$, $b = 0$ as in Definition 9

and by obtaining the expansion of $x(\epsilon) = \|\mathrm{Log}_\mathrm{I}\{\mu(D'_\epsilon)\}\|$ in both cases.

By [20] for function $f(y) = \mathrm{d}(y,p)^2 = \|\mathrm{Log}_y(p)\|^2$ we have

$$\nabla_y f = -2\mathrm{Log}_y(p_i). \tag{31}$$

Using this gradient in Definition (5) $x(\epsilon)$ is such that

$$\sum_{i=1}^N \mathrm{Log}_\mathrm{I}[\mathrm{Exp}_\mathrm{I}\{-x(\epsilon)/2\}\mathrm{Exp}_\mathrm{I}(\epsilon r_i)\mathrm{Exp}_\mathrm{I}\{-x(\epsilon)/2\}] = 0 \tag{32}$$

with

1. $\mathrm{Exp}_\mathrm{I}(\epsilon r_i) = \mathrm{Exp}_\mathrm{I}\{-t_i(\epsilon, v_k)v_k/2\}\mathrm{Exp}_\mathrm{I}(\epsilon q_i)\mathrm{Exp}_\mathrm{I}\{-t_i(\epsilon, v_k)v_k/2\}$ and

2. $\mathrm{Exp}_\mathrm{I}(\epsilon r_i) = \mathrm{Exp}_\mathrm{I}\{-t_i(\epsilon, v_k)v_k\}\mathrm{Exp}_\mathrm{I}(\epsilon q_i)$

for $i = 1, \ldots, N$ in cases 1 and 2, respectively. Letting

$$x(\epsilon) = x_1\epsilon + x_2\epsilon^2 + x_3\epsilon^3 + x_4\epsilon^4 + O(\epsilon^5)$$

substituting into (32) and solving for $x_1, x_2, x_3$ and $x_4$ gives the expansions

1. $x(\epsilon) = x_3 \epsilon^3 + O(\epsilon^5)$ where

$$\begin{aligned}
x_3 &= \frac{1}{96} \sum_{i=1}^{N} \{\operatorname{tr}(q_i v_k)^2\}(2v_k q_i v_k - q_i v_k v_k - v_k v_k q_i) - \\
&\qquad 4\{\operatorname{tr}(q_i v_k)\}(2q_i v_k q_i - q_i q_i v_k - v_k q_i q_i) \\
&\qquad + 4\{\operatorname{tr}(q_i v_k)\}\{\operatorname{tr}(q_i q_i v_k v_k) - \operatorname{tr}(q_i v_k q_i v_k)\} v_k \\
&= \frac{1}{24} \sum_{i=1}^{N} \|q_i\|^3 \Big[ \cos^2 \theta_i R(q_i, v_k) v_k + 2\cos \theta_i R(v_k, q_i) q_i \\
&\qquad - 2\cos \theta_i \sin^2 \theta_i K(\sigma_{q_i, v_k}) v_k \Big] \text{ and}
\end{aligned}$$

2. $x(\epsilon) = x_2 \epsilon^2 + O(\epsilon^3)$ where

$$x_2 = \sum_{i=1}^{N} (1/4) \operatorname{tr}(q_i v_k)(v_k q_i - q_i v_k) = -\sum_{i=1}^{N} (1/2) \|q_i\|^2 \cos \theta_i [v_k, \tilde{q}_i]$$

in case 1 and case 2, respectively with $\|x(\epsilon)\| = O(\epsilon^3)$ in case 1 and $\|x(\epsilon)\| = O(\epsilon^2)$ in case 2.

Considering these expansions, $a = b = 1/2$ gives less local displacement of the mean and $a = b = 1/2$ is preferable. Still, the use of a centering or normalization step in this form of PGA might be considered, particularly when there is still significant variability in the data and if some degree of degeneracy in explanatory directions is observed.

In table 2 in 500 runs I generated 50 vectors in $T_I SO(3)$ with entries having a standard Gaussian distribution and differing variances. I compare the mean angles over the 500 runs that located explanatory directions, both for $a = 1, b = 0$ (as in Definition 9) and $a = b = 1/2$, make with the eigenvectors of covariance operator $L$ ($\theta$ w/ e.v.'s) and with PGA directions in Definition 7 ($\theta$ w/ PGA). I also measure the displacement of the intrinsic mean after removing an explanatory direction ($\mu$ disp.) and the *average reconstruction error*, i.e., the intrinsic variability remaining in the data after removing explanatory directions (R.E.) under $a = b = 1/2$ and $a = 1, b = 0$.

We repeat the experiment for data given in [26] which was collected to investigate the variability in six subjects' movements while completing a drilling task. The data we use is motion capture data of the rotation of the first subject's wrist. This data has little variability which is claimed in [26] is common in human kinetics studies and accordingly in [26] tangent space methods are used for analysis.

The tables show better agreement with the eigenvectors of L and variability as identified by PGA for $a = b = 1/2$. Also, there is less displacement of the mean for $a = b = 1/2$ in agreement with the expansions above. There is also slightly less reconstruction error for $a = b = 1/2$ in these experiments. These effects are greater in magnitude for the simulated data which has greater tangent space variability.

|  |  | $\theta$ w/ e.v.'s | $\theta$ w/ PGA | $\mu$ disp. | R.E. |
|---|---|---|---|---|---|
| $a = b = \frac{1}{2}$ | first dir. | 0.0844 | 0.00 | 0.0109 | 0.4640 |
|  | second dir. | 0.1180 | 0.0021 | 0.0057 | 0.2530 |
| $a = 1, b = 0$ | first dir. | 0.0844 | 0.00 | 0.0196 | 0.4640 |
|  | second dir. | 0.6322 | 0.5880 | 0.0112 | 0.2610 |

Simulated data, 500 runs, mean intrinsic var. = 1.4878

|  |  | $\theta$ w/ e.v.'s | $\theta$ w/ PGA | $\mu$ disp. | R.E. |
|---|---|---|---|---|---|
| $a = b = \frac{1}{2}$ | first dir. | 1.308e-3 | 0.00 | 110.418e-6 | 43.853e-3 |
|  | second dir. | 2.01e-3 | 27.997e-6 | 42.035e-6 | 19.926e-3 |
| $a = 1, b = 0$ | first dir. | 1.308e-3 | 0.00 | 437.222e-6 | 43.853e-3 |
|  | second dir. | 63.526e-3 | 61.940e-3 | 189.840e-6 | 20.247e-3 |

Wrist rotation data, intrinsic var. = 0.2912

Table 2: Comparisons of alt-PGA

## 7. Linear Difference Indicators

In [28] differences between exact solutions to PGA and tangent space approximations were explored. To this end [28] introduced measures of the accuracy of approximations of the projection operator and of approximations of explanatory directions obtained by orthogonal projection and PCA in the tangent space, respectively. Given $D = \{p_1, \ldots, p_N\} \subset M$ with $\mu(D) = \mu$, PGA directions $V_{k-1} = \{v_1, \ldots, v_{k-1}\}$ and $v \in SV_{k-1}^\perp$ the *average projection difference* is formulated as

$$\tau_H = \frac{1}{N} \sum_{i=1}^{N} \mathrm{d}\{p_i, \hat{\pi}_{H(v)}(p_i)\}^2 - \mathrm{d}\{p_i, \pi_{H(v)}(p_i)\}^2 \qquad (33)$$

where $H(v) = \mathrm{Exp}_\mu(\mathrm{span}(V_{k-1} \cup v))$ and $\hat{\pi}_{H(v)}(p_i)$ is the exponentiation of the orthogonal projection of $q_i = \mathrm{Log}_\mu(p_i)$ in $\mathrm{T}_\mu M$ to $\mathrm{span}(V_{k-1} \cup v)$.

Then setting $v = v_k$ where $v_k$ is the $k$th PGA direction and letting $\hat{v}$ be its first-order approximation obtained in $\mathrm{T}_\mu M$ the *average residual difference* is formulated as

$$\rho = \frac{1}{N} \sum_{i=1}^{N} \mathrm{d}\{p_i, \pi_{H(\hat{v})}(p_i)\}^2 - \mathrm{d}\{p_i, \pi_{H(v)}(p_i)\}^2$$

In [28] *difference indicators* which are shown to be correlated to these statistics and which can be computed before carrying out exact PGA are proposed. The indicator for the average projection difference is given as

$$\tau_H \approx \tilde{\tau}_{H(v)} = \frac{2}{N} \sum_{i=1}^{N} \|\nabla_{\hat{\pi}_{H(v)}(p_i)} f\|, \qquad (34)$$

where $f(y) = d(p_i, y)^2$. Using (31), $\|\nabla_{\hat{\pi}_{H(v)}(p_i)} f\|$ can be computed as the magnitude of the component of $-2\text{Log}_{\hat{\pi}_{H_v}(p_i)}(p_i)$ in $T_{\hat{\pi}_{H(v)}(p_i)} M$.

The indicator for the average residual difference is given as the standard deviation of the differences of the distances of the $q_i's$ to their orthogonal projections to $\text{span}(V_{k-1} \cup \hat{v})$ and the distances of the $p_i's$ to the exponentiation of those orthogonal projections. That is,

$$\sigma = \sqrt{\frac{1}{N} \sum_{i=1}^{N} \left[\|q_i - \text{Log}_\mu\{\hat{\pi}_{H(\hat{v})}(p_i)\}\| - d\{p_i, \hat{\pi}_{H(\hat{v})}(p_i)\} - m\right]^2}$$

where $m$ is the mean of the differences between the distances.

### 7.1. Expansions of $\tau_H$ and $\rho$

We obtain expansions in scaling parameter $\epsilon$ of $\tau_H$ and $\rho$ in $P(n)$ and $SO(n)$. Then in Section 7.2 we apply the difference indicators and the expansions to two data sets in experiments similar to the ones in [28] and show the expansions provide good approximations of $\tau_H$ and $\rho$. Note that although $\tilde{\tau}_{H(v)}$ and $\sigma$ show correlation only $\tilde{\tau}_{H(v)}$ is an approximation. These expansions will also give a better understanding of how scale and curvature terms affect $\tau_H$ and $\rho$ which then can be used to make the type of decisions about the use of linear approximations demonstrated in [28]. We will obtain these expansions for $k = 1$ which will allow us to carry out the type of experiments done in [28]. Expansions for $k > 1$ can be obtained in a similar manner.

Let $p_{i,\epsilon} = \text{Exp}(\epsilon q_i)$ and $t_i(\epsilon) = t_{i,1}\epsilon + t_{i,3}\epsilon^3 + O(\epsilon^5)$ be the expansion of the projection coefficient to a geodesic as Section 5.3. Using the expansion of the objective function in (24) and of $t_i(\epsilon)$ given in (25) we obtain

$$\tau_H = \frac{1}{N} \sum_{i=1}^{N} h_i(t_{i,1}, \epsilon) - h_i(t_i(\epsilon), \epsilon)$$

$$= \frac{1}{N} \sum_{i=1}^{N} (t_{i,3})^2 \epsilon^6 + O(\epsilon^8)$$

$$= \frac{1}{144N} \sum_{i=1}^{N} \|q\|^6 \cos^2(\theta_i) \sin^4(\theta_i) \{K(\theta_{v,q_i})\}^2 \epsilon^6 + O(\epsilon^8)$$

Consider the cost function $f_1(v, \epsilon)$ in (2) and the expansion

$$v_1(\epsilon) = v_{1,0} + v_{1,2}\epsilon^2 + v_{1,4}\epsilon^4 + O(\epsilon^6).$$

We have
$$\rho = f_1(v_{1,0}, \epsilon) - f_1\{v_1(\epsilon), \epsilon\}. \tag{35}$$

Given the constraint $\|v_1(\epsilon)\| = 1$ and that for

$$v_1^a(\epsilon) = \sum_{j=0}^{a} v_{2j}\epsilon^{2j},$$

$\|v_1^a(\epsilon)\|$ is not necessarily 1 for any $a$, consider the expansion

$$\frac{1}{\|v_1^a(\epsilon)\|} = \frac{1}{\sqrt{1 + \langle v_{1,2}, v_{1,2}\rangle_\mathrm{I}\epsilon^4 + O(\epsilon^8)}} = 1 - \frac{\langle v_{1,2}, v_{1,2}\rangle_\mathrm{I}}{2}\epsilon^4 + O(\epsilon^8)$$

and set

$$\begin{aligned}\tilde{v}_1^a(\epsilon) &= v_1^a(\epsilon)/\|v_a(\epsilon)\| \\ &= v_{1,0} + v_{1,2}\epsilon^2 + \left(v_{1,4} - v_{1,0}\frac{\langle v_{1,2}, v_{1,2}\rangle_\mathrm{I}}{2}\right)\epsilon^4 + O(\epsilon^6).\end{aligned}$$

Then we have $v_1(\epsilon) = \lim_{a\to\infty}\tilde{v}_1^a(\epsilon)$ and as in (5) we have

$$\begin{aligned}[f_{1,2}(v_{1,0}) &- f_{1,2}\{v_1(\epsilon)\}]\epsilon^2 \\ &= \left(\frac{1}{N}\sum_{i=1}^n \langle q_i, v_1(\epsilon)\rangle_\mathrm{I}^2 - \langle q_i, v_{1,0}\rangle_\mathrm{I}^2\right)\epsilon^2 \\ &= \frac{1}{N}\sum_{i=1}^N \left(\langle q_i, v_{1,2}\rangle_\mathrm{I}^2 - \langle q_i, v_{1,0}\rangle_\mathrm{I}^2\langle v_{1,2}, v_{1,2}\rangle_\mathrm{I}\right)\epsilon^6 + O(\epsilon^8)\end{aligned} \quad (36)$$

with the second equality using

$$\frac{1}{N}\sum_{i=1}^N \langle q_i, v_{1,0}\rangle_\mathrm{I}\langle q_i, v_{1,k}\rangle_\mathrm{I} = \langle \mathrm{L}(v_{1,0}), v_{1,k}\rangle_\mathrm{I} = \beta_1\langle u_1, v_{1,k}\rangle_\mathrm{I} = 0$$

for $k = 2, 4$. Using (28), in $P(n)$ we have

$$\begin{aligned}[f_{1,4}&(v_{1,0}) - f_{1,4}\{v_1(\epsilon)\}]\epsilon^4 \\ &= \frac{1}{48N}\sum_{i=1}^N \Big\{\mathrm{tr}(q_i v_{1,0})^2\mathrm{tr}\left(2q_i v_{1,0}q_i v_{1,2} - q_i q_i v_{1,0}v_{1,2} - q_i q_i v_{1,2}v_{1,0}\right) \\ &\quad + 2\mathrm{tr}\left(q_i v_{1,0}\right)\mathrm{tr}\left(q_i v_{1,2}\right)\mathrm{tr}\left(q_i v_{1,0}q_i v_{1,0} - q_i q_i v_{1,0}v_{1,0}\right)\Big\}\epsilon^6 + O(\epsilon^8) \\ &= \frac{1}{6N}\sum_{i=1}^N \Big\{\cos^2\theta_{i,1}\mathrm{R}_\mathrm{I}(v_{1,0}, \tilde{q}_i, \tilde{q}_i, v_{1,2}) \\ &\quad + \cos\theta_{i,1}\sin^2\theta_{i,1}\cos\hat{\theta}_{i,1}K(\sigma_{v,q_i})\Big\}\|q_i\|^6\epsilon^6 + O(\epsilon^8).\end{aligned} \quad (37)$$

In $SO(n)$, using the metric in (30), we just take the negative of this expression. Then using (35) the expansion of $\rho$ in $P(n)$ or $SO(n)$ is the sum of (36) and (37).

*7.2. Experiments comparing difference indicators and expansions*

Set the expansion of $\rho$ as $\rho = \rho_6\epsilon^6 + O(\epsilon^8)$ and the expansion of $\tau_H$ as $\tau_H = \tau_{H,6}\epsilon^6 + O(\epsilon^8)$, as derived above in Section 7.1. The first data set is the wrist rotation data set in $SO(3)$ from Section 6.3. The second is a synthetic data set in $P(n)$ with a sample of 36 from a distribution as in the test in Figure 6. As in [28] for each experiment we draw a random sample of size 8 from the data set 20 times and compute the relevant statistics each time for comparison. For the wrist rotation data in $SO(3)$, $\tau_H$ is not computed as projection has closed-form in this case.

**Experiment 1. Wrist Rotation Data**

As indicated in Section 6.3 this data set has little variability. Accordingly, $\rho_6$ in Figure 7 is a very close estimate of $\rho$ and provides a nearly perfect picture of the penalty (very small in this case) of using the first-order approximation of PGA. At the same time $\sigma$ has a lower correlation with $\rho$ and is of a much different scale and is thus of less value in assessing the use of a first-order approximation of PGA.

**Experiment 2. Synthetic Data**

As in [28] the first PGA direction $v_1$ is set to $v$ in (33) and used to compute $\tau_H$. In Figure 8 both $\tau_{H,6}$ and $\rho_6$ provide good estimates of $\tau_H$ and $\rho$ with correlation coefficients of .993 and .981, respectively. Also to note, one might *square* the norm of the gradient in (34) to use the Pythagorean theorem in each $T_{\hat{\pi}_{H(v)}(p_i)}M$ to get another improved estimate of $\tau_H$ over $\tilde{\tau}_H$.

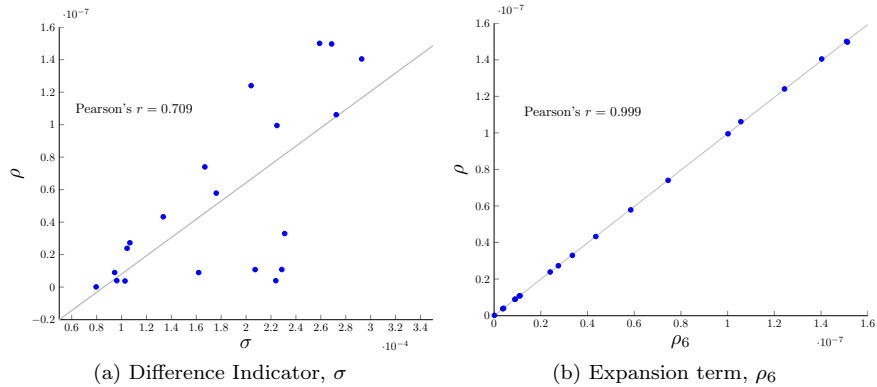

(a) Difference Indicator, $\sigma$     (b) Expansion term, $\rho_6$

Figure 7: $\rho$ for wrist rotation

## 8. Discussion and conclusions

In [31] PGA is formulated as a probability model (PPGA) in which data is distributed according to a manifold generalization of the normal distribution. Explanatory directions are included as parameters to be estimated by maximum likelihood. Also, a location parameter and scaling parameters for the explanatory directions and for the variability or dispersion of the data are fit. Thus, as an advantage, not only are the the mean and explanatory directions jointly estimated but the dispersion of the data is also taken into account.

In the descriptive setting, in this paper, consideration of a dispersion or scaling factor was shown to be an essential element in revealing the underlying structure of solutions to PGA. In the proposition in Section 3, for example, we see how the share of variability in the tangent space, accounted for by eigenvector $u_k$ and measured by eigenvalue $\beta_k$, weights the curvature terms in $v_{k,2}$ to determine the difference, at least locally, between the first-order approximation and the exact solution. Or in expansions of Section 7 we see the explicit inter-action between scale, curvature and the distribution of data in determining the local difference between the projection operator, the PGA directions and their linear approximations.

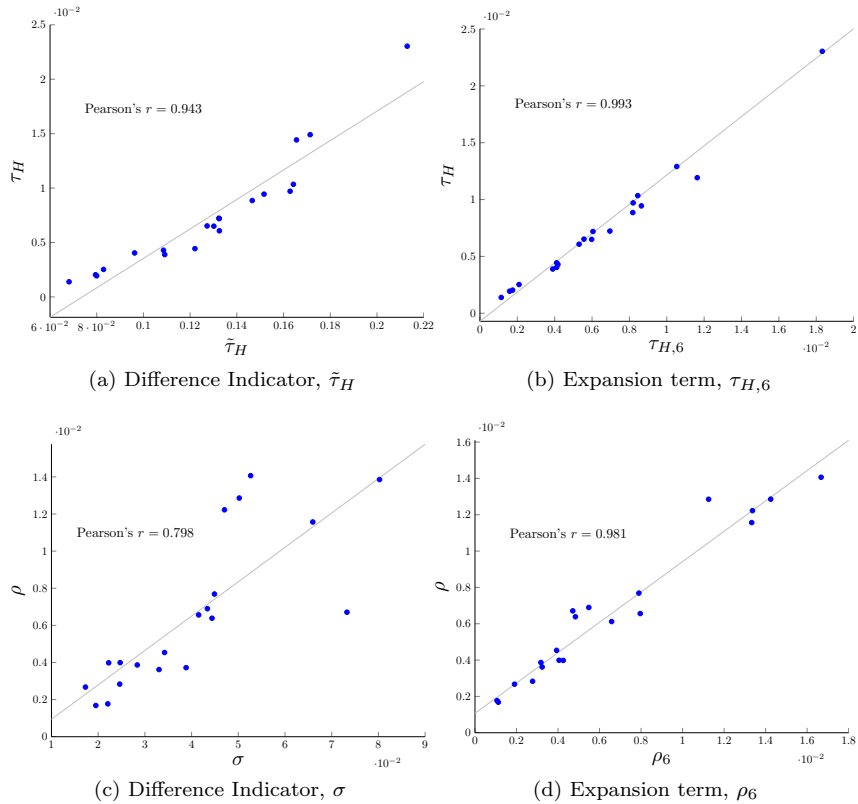

(a) Difference Indicator, $\tilde{\tau}_H$  (b) Expansion term, $\tau_{H,6}$

(c) Difference Indicator, $\sigma$  (d) Expansion term, $\rho_6$

Figure 8: $\tau_H$, $\rho$ for data in $P(3)$

Also, we see in this paper, at least experimentally, that the approximations obtained by expansion hold for data significantly dispersed from the tangent space. In Figure 4, for example, with $q, v$ uniformly distributed in $S_I P(n)$ the third- and fifth-order approximations in $\epsilon$ hold for $\epsilon > 2$. In this case, one can readily obtain bounds on the ratios of the coefficients in the expansion and derive expected values of those ratios to explain this plot. Such an approach should be able to be taken with other expansions and distributions of data as well. Also, in a bounded manifold the variability of data and thus $\epsilon$ is restricted. In $S^n$ or $SO(n)$, for example, we take $|\epsilon| < \pi / \|q_i\|$ for data $p_{i,\epsilon} = \text{Exp}_\mu(\epsilon q_i)$ so that the expansions only need to hold for these values of $\epsilon$.

Since PGA was introduced in [11] and then [10] a number of other methods to analyze the variability of manifold-valued data have been proposed. For example, [18] accounts for non-geodesic variability in spheres and [17] projects to geodesics that intersect orthogonally at a mean on the first geodesic that best fits the data. One could use the approach of this paper in these contexts, for example, an expansion of the difference between the mean located in [17] and the intrinsic mean might be obtained.

In addition, this paper and others use a definition of PGA that minimizes residual error while PGA was defined in [11] as maximizing projected variability. Expansions of solutions of PGA using the latter definition might be obtained and the higher-order terms compared to determine what accounts for the differences and how these differences might be taken into account in deciding which definition to employ. Also, the approach of this paper might be used to quantify differences between other generalizations of linear statistics such as *intrinsic MANOVA* in [15] and *geodesic regression* in [9] and their local, linear approximations.

## 9. Acknowledgments

We are grateful to the Editor, the Associate Editor, and the reviewers for their valuable comments which have led to improvements of our paper. DL thanks Professor Leo Butler for help and guidance in working on this paper. We also thank Professor Steve Rosenberg for useful discussions. The contribution of LL was funded by NSF grant IIS1546331 and a grant from the Army's research office.

## References

[1] A. Banerjee, I. S. Dhillon, J. Ghosh, and S. Sra, *Clustering on the unit hypersphere using von Mises–Fisher distributions*, J. Mach. Learn. Res. **6** (2005), 1345–1382.

[2] P.J. Basser, J. Mattiello, and D. LeBihan, *{MR} diffusion tensor spectroscopy and imaging*, Biophysical Journal **66** (1994), 259 – 267.

[3] M. Bridson and A. Haefliger, *Metric spaces of non-positive curvature*, Grundlehren der mathematischen Wissenschaften, 319, pp. 315–316, Springer, 1999.

[4] J. Cheeger and D.G. Ebin, *Comparison theorems in Riemannian geometry*, AMS Chelsea Publishing, Providence, RI, 2008, Revised reprint of the 1975 original.

[5] R.W.R. Darling, *Differential forms and connections*, Cambridge University Press, Cambridge, 1994.

[6] M.P. do Carmo, *Differential geometry of curves and surfaces*, Prentice-Hall, Inc., Englewood Cliffs, N.J., 1976, Translated from the Portuguese.

[7] P. Jost-Hinrich Eschenburg, *Lecture notes on symmetric spaces*, January 1997, http://myweb.rz.uni-augsburg.de/ eschenbu/symspace.pdf.

[8] P.T. Fletcher, *Lecture notes on symmetric spaces*, January 2009, http://www.sci.utah.edu/fletcher/RiemannianGeometryNotes.pdf.

[9] \_\_\_\_\_\_, *Geodesic regression and the theory of least squares on Riemannian manifolds*, Int. J. Comput. Vis. **105** (2013), 171–185.

[10] P.T. Fletcher and S. Joshi, *Principal geodesic analysis on symmetric spaces: Statistics of diffusion tensors*, Computer Vision and Mathematical Methods in Medical and Biomedical Image Analysis (Milan Sonka, IoannisA. Kakadiaris, and Jan Kybic, eds.), Lecture Notes in Computer Science, vol. 3117, Springer Berlin Heidelberg, 2004, pp. 87–98 (English).

[11] P.T. Fletcher, Conglin Lu, and Sarang Joshi, *Statistics of shape via principal geodesic analysis on lie groups*, Proceedings of the 2003 IEEE Computer Society Conference on Computer Vision and Pattern Recognition (Washington, DC, USA), CVPR'03, IEEE Computer Society, 2003, pp. 95–101.

[12] M. Fréchet, *Les élements aléatoires de nature quelconque dans un espace distancié*, Ann. Inst. H. Poincaré **10** (1948), 215–310.

[13] J. Gallier and D. Xu, *Computing exponentials of skew symmetric matrices and logarithms of orthogonal matrices*, International Journal of Robotics and Automation **18** (2000), 10–20.

[14] K. Hornik, I. Feinerer, M. Kober, and C. Buchta, *Spherical k-Means Clustering*, Journal of Statistical Software **50** (2012), 1–22.

[15] S. Huckemann, T. Hotz, and A. Munk, *Intrinsic manova for riemannian manifolds with an application to kendall's space of planar shapes.*, IEEE Trans. Pattern Anal. Mach. Intell. **32** (2010), 593–603.

[16] ______ , *Intrinsic shape analysis: geodesic PCA for Riemannian manifolds modulo isometric Lie group actions*, Statist. Sinica **20** (2010), 1–58.

[17] S. Huckemann and H. Ziezold, *Principal component analysis for Riemannian manifolds, with an application to triangular shape spaces*, Adv. in Appl. Probab. **38** (2006), 299–319.

[18] S. Jung, I.L. Dryden, and J.S. Marron, *Analysis of principal nested spheres*, Biometrika **99** (2012), 551–568.

[19] B.K. and P. Horn, *Relative orientation*, International Journal of Computer Vision (1990), 59–78.

[20] H. Karcher, *Riemannian center of mass and mollifier smoothing*, Comm. Pure Appl. Math. **30** (1977), 509–541.

[21] J. Kuffner, *Effective sampling and distance metrics for 3d rigid body path planning*, Robotics and Automation, 2004. Proceedings. ICRA '04. 2004 IEEE International Conference on, vol. 4, April 2004, pp. 3993–3998.

[22] K. Lange, *Optimization*, Springer Texts in Statistics, 95, p. 36, Springer, 2013.

[23] K.V. Mardia and P.E. Jupp, *Directional statistics*, Wiley Series in Probability and Statistics, John Wiley & Sons, Ltd., Chichester, 2000, Revised reprint of Statistics of directional data by Mardia.

[24] I. Najfeld and T.F. Havel, *Derivatives of the matrix exponential and their computation*, Adv. in Appl. Math. **16** (1995), 321–375.

[25] P. Petersen, *Riemannian geometry*, Graduate Texts in Mathematics, p. 242, Springer, 2010.

[26] D. Rancourt, L.-P. Rivest, and J. Asselin, *Using orientation statistics to investigate variations in human kinematics*, J. Roy. Statist. Soc. Ser. C **49** (2000), 81–94.

[27] S. Said, N. Courty, N. Le Bihan, and S.J. Sangwine, *Exact principal geodesic analysis for data on SO(3)*, Signal Processing Conference, 2007 15th European, Sept 2007, pp. 1701–1705.

[28] S. Sommer, F. Lauze, S. Hauberg, and Mads Nielsen, *Manifold valued statistics, exact principal geodesic analysis and the effect of linear approximations*, Proceedings of the 11th European Conference on Computer Vision: Part VI (Berlin, Heidelberg), ECCV'10, Springer-Verlag, 2010, pp. 43–56.

[29] S. Sommer, F. Lauze, and M. Nielsen, *The differential of the exponential map, jacobi fields and exact principal geodesic analysis*, CoRR **abs/1008.1902** (2010), 1–26.

[30] B. Stanfill, H. Hofmann, and U. Genschel, *Rotations: An R package for SO(3) data*, The R Journal **6** (2014), 68–78.

[31] M. Zhang and P.T. Fletcher, *Probabilistic principal geodesic analysis.*, NIPS (Christopher J. C. Burges, Lon Bottou, Zoubin Ghahramani, and Kilian Q. Weinberger, eds.), 2013, pp. 1178–1186.